# Study of spin glass and cluster ferromagnetism in $RuSr_2Eu_{1.4}Ce_{0.6}Cu_2O_{10-\delta}$ magneto Superconductor

Anuj Kumar[1,2], R.P.Tandon[2] and V.P.S. Awana[1,*]

[1]Quantum Phenomena and Application Division, National Physical Laboratory (CSIR)

Dr. K. S. Krishnan Road, New Delhi-110012, India

[2]Department of Physics and Astrophysics, University of Delhi, New Delhi-110007, India

We report *DC* magnetization, detailed systematic linear and nonlinear *AC* magnetic susceptibility and transport for a single phase $RuSr_2Eu_{1.4}Ce_{0.6}Cu_2O_{10-\delta}$ (EuRu-1222) magneto-superconductor. The studied sample is synthesized through standard solid state reaction route, which is crystallized in single phase tetragonal structure with space group *I4/mmm*. *DC* magnetic susceptibility measurements revealed that the studied EuRu-1222 is a magneto-superconductor with Ru spins ordering at around 110 K and superconductivity in the Cu-$O_2$ planes below ≈ 30 K. Temperature dependence of *AC* susceptibility with different frequency and amplitude variations confirms spin-glass behavior with cluster ferromagnetism of the system. Change in the cusp position with frequency follows the *Vogel-Fulcher law*, which is commonly accepted feature for a spin-glass (SG) system with ferromagnetic clusters. The third harmonic of *AC* susceptibility ($\chi_3$) shows that the system undergoes a spin glass transition below 80 K. Superconducting transition temperature ($T_c$) onset and $\rho = 0$ are seen at around 30 and 18 K without any applied field and the same decreases to 10 and 2 K under 130 kOe applied field. Also low fields isothermal *(MH)* suggests that ferromagnetic clusters are embedded in spin-glass (SG) matrix. The magnetization vs. applied field (*MH*) loops exhibited ferromagnetic (*FM*) like behavior with rather small coercive fields. Detailed *AC* magnetic susceptibility measurements are carried out to unearth the short range magnetic correlations. These results support the spin-glass (SG) formation followed by ferromagnetic clustering effects at low temperatures. Our detailed magnetization and magneto transport results will undoubtedly contribute to current understanding of the complex magnetism of the EuRu-1222 system.

***Corresponding author:** awana@mail.nplindia.ernet.in/(www.freewebs.com/vpsawana)

**Tel.** +91 11 45609357, **Fax.** +91 11 45609310

## Introduction

The magnetic properties of rutheno-cuprates, which belongs to the class of high temperature superconductors (*HTSc*), had raised significant interest in scientific community due to the presence of Ru ion magnetic order being co-existing with superconductivity. Co-existence of superconductivity and magnetic order in the hybrid rutheno-cuprates $RuSr_2(Eu,Gd,Sm)_{1.5}Ce_{0.5}Cu_2O_{10-\delta}$ (Ru-1222) and $RuSr_2(Eu,Gd,Sm)Cu_2O_{8-\delta}$ (Ru-1212) is interesting because the magnetic ordering temperature is much higher (~150 K) than the superconducting transition temperature (~30 K) and both do co-exist in the same unit cell of these systems [1,2]. A review of some earlier magnetic superconductors including $RRh_4B_4$ and $RMo_6S_8$ compounds (R = rare earth) is presented in ref. [3]. Rutheno-cuprates appear to be the most challenging systems in term of understanding the novel phenomena of co-existing superconductivity and ferromagnetism. More than a decade is passed and scientists are yet trying to understand complex magnetism of these systems [4-6]. The magnetic behavior of Ru-1222 is even more complicated than Ru-1212 phase. The first simultaneous observation of superconductivity and the Ru ion magnetic order in a rutheno-cuprate was published in 1997 for $RuSr_2RE_{1.4}Ce_{0.6}Cu_2O_{10-\delta}$ (here RE represents the rare earth ion Eu or Gd) [1]. This is the so called Ru-1222 type phase (the numbers reflect the number of metal atoms present in their distinct crystallographic positions per formula, sometimes also named the 2122-type, with RE site occupancy being listed first). Muon spin rotation (*μSR*) experiments were performed on $RuSr_2GdCu_2O_{8-\delta}$ (Ru1212) type phase and the presence of bulk magnetic order $T_m$ = 133 K was confirmed with superconductivity at below 16 K [2, 4]. The magnetic ordering of transition metal magnetism ($T_m$) sets above the superconducting transition temperature ($T_c$) (e.g. for $RuSr_2GdCu_2O_{8-\delta}$: $T_m \approx 132$ K and $T_c \approx 45$ K [5, 6]). While the co-existence of superconductivity with low temperature antiferromagnetic (*AFM*) ordering in some of the rare earth ions (e.g. for $GdBa_2Cu_3O_{7-\delta}$, $T_c$ = 92 K and $T_N$ = 2.2 K [7]) is well known and has been investigated, the $T_m > T_c$ is observed only for rutheno-cuprates system. Ferromagnetic-like and superconducting properties have also been reported for $RuSr_2EuCu_2O_8$ ($T_m$ = 132 K, $T_c \leq 25$ K) [2, 8], $RuSr_2YCu_2O_8$ ($T_m$ = 149 K, $T_c \leq 39$ K) [9] and $RuSr_2RECu_2O_8$ for RE = Dy, Ho, Er [10, 11]. A phase separation into ferromagnetic (*FM*) clusters and paramagnetic (*PM*) matrix followed by an antiferromagnetic (*AFM*) transition has been assumed to establish magnetic correlations in these systems [12]. Neutron diffraction data [13] did not show unique long range magnetic order, which is encountered more recently [14], primarily in Ru-1222 phase. These results are yet debated in literature. Also, interacting magnetic clusters in the Ru-$O_2$ planes with no long-range magnetic order, was proposed recently in $Nb_{1-x}Ru_x$-1222 system [15]. The slow spin dynamics suggested that the *FM* clusters in Ru-1222 could exhibit superparamagnetism (*SPM*) [12]. An interesting observation of the frequency dependent peak shift of the *AC* susceptibility as a function of temperature with thermoremanent magnetization (*TRM*) measurements [16] revealed spin-glass (*SG*) behavior in Ru1222, which in a way contradicted the existence of long range magnetic order in these systems. Spin glasses are magnetic systems in which the magnetic moments are ‘‘in conflict’’ with each other due to some frozen-in structural disorder. Thus no convectional long-range order (of ferromagnetic or antiferromagnetic type) could be established. These systems exhibit a ‘‘freezing transition’’ to a state of new kind of order in which spins are aligned in random directions [17]. Spin-glass (SG) is simply the kinetic arrest of some sort of ferromagnetic and antiferromagnetic spins. One of the characteristic phenomena observed in spin glasses systems is the occurrence of sharp cusp in the frequency dependent *AC* susceptibility in low fields. The physics of spin glass raises many fundamental questions and has become one of the main streams of research particularly in condensed matter physics. A frequency and field dependent sharp cusp is observed in *AC* susceptibility around 100 K, indicating metastable magnetism in these systems [12, 15, 16]. These experimental features are typically found in spin-glass (*SG*), cluster-glass (*CG*) or superparamagnets (*SPM*) and even in inhomogeneous ferromagnets etc. [17]. Superparamagnetism (*SPM*) is a dynamic phenomenon, which may exhibit paramagnetism, even below the Curie (or the Neel) temperatures [17].

One of the challenging issues with these compounds is their synthesis in single phase and understanding of their complex magnetism. Our aim is to probe the complex magnetic nature of this compound using bulk susceptibility measurements. We concentrated on some aspects, which we feel are important to unearth the complex magnetism of these systems. *AC* susceptibility and its various harmonic studies is one of the best tools to analyze such type of magnetically complicated systems. The magnetization measured in the presence of an excitation field (*AC* field) can be written as power law [17]

$$M = M_0 + \chi_1 H + \chi_2 H^2 + \chi_3 H^3 + \chi_4 H^4 + \ldots\ldots\ldots\ldots \quad (1)$$

Where $\chi_1$, $\chi_2$, $\chi_3$ are the first, second, third order harmonics of the *AC* susceptibility respectively [17, 25], providing useful information about the dynamics of existing/competing magnetic order.

Although, frequency dependence of the peak and slow magnetic relaxation are generic features of both *SG* and *SPM*, the physics behind them is different. Thus, a more careful investigation of the EuRu-1222 system is required to probe the existence of these two metastable states. The existence of the glassy state in the vicinity of the *FM* transition gives rise to two possible types of *SG* behavior: reentrant-spin-glass (*RSG*) [17] or cluster-glass (*CG*) [12]. In *RSG* systems a *PM* to *FM* transition is seen as a bifurcation in zero-field cooled (*ZFC*) and field cooled (*FC*) curves. While for CG the frequency and applied *AC* field dependence of the peak is essential feature, although *the exact shift is different*.

Keeping in view the complex magnetism of EuRu-1222, we report in this article detailed magnetization (*DC*/*AC*) and magneto-transport of a single phase EuRu-1222 compound namely $RuSr_2Eu_{1.4}Ce_{0.6}Cu_2O_{10-\delta}$. Our results support the spin-glass formation followed by ferromagnetic clustering effects at low temperatures.

## Experimental details

Polycrystalline bulk sample of $RuSr_2Eu_{1.4}Ce_{0.6}Cu_2O_{10-\delta}$ (EuRu-1222) was synthesized through solid state reaction route from stoichiometric powders with 99.99% purity $RuO_2$, $SrCO_3$, $Eu_2O_3$, $CeO_2$ and CuO. These mixtures were ground together in an agate mortar and calcined in air at $1020^o$ C, $1040^o$ C and $1060^o$ C each for 24 hours with intermediate grindings. The pressed bar shaped pellet was annealed in Oxygen atmosphere at $850^o$ C, $650^o$ C and $450^o$ C each for 24 hours, and subsequently cooled down slowly over a span of 12 hours to the room temperature. X-ray diffraction (*XRD*) was performed at room temperature in the scattering angular (*2θ*) range of $20^o$-$80^o$ in equal steps of $0.02^o$ using *Rigaku* diffractometer with *Cu* $K_\alpha$ (λ = 1.54Å) radiation. Rietveld analysis was performed using the *FullProf* program. Detailed *DC* and *AC* (linear and nonlinear) magnetization were performed on Physical Property Measurements System (*PPMS*-14T, Quantum Design-USA) as a function of temperature and applied magnetic field both. Linear and nonlinear *AC* susceptibilities as a function of temperature (*T*), (i) in the frequency ranges of 33-9999 *Hz* and, (ii) in the drive *AC* magnetic field amplitude of 1-17 *Oe* in zero external *DC* magnetic fields, were also measured on *PPMS*. Electrical resistivity (*ρ*) measurement over the temperature range of 1.9 K to 250 K, in various fields of maximum up to 13 kOe using the simple four probe technique was also performed on *PPMS*.

## Results and Discussion

The quality of sample in case of rutheno-cuprates is important for any meaningful discussion, since minute impurities of $SrRuO_3$ and $Sr_2EuRuO_6$ ($211O_6$) phases tend to form readily in the matrix. Any small impurity of these foreign phases may complicate the resultant outcome magnetization. After several calcinations and grinding processes as mentioned above, we finally obtained a single phase EuRu-1222 compound. It was observed that characteristic peaks corresponding to $SrRuO_3$ and $Sr_2EuRuO_6$ phases are not observed within the *X*-ray detection limit. This is very important because both of these phases, which generally develop during the synthesis of rutheno-cuprates, are magnetic in nature, and hence they significantly alter the net outcome magnetization. Observed and fitted X-ray pattern for the compound is shown in Figure 1. The structure analysis was performed using the Rietveld refinement analysis by using the FullProf Program. The Rietveld analysis confirms a tetragonal single phase formation in *I4/mmm* space group. All the Reitveld refined structural parameters (lattice parameters, site occupancy and atomic coordinates) of compound are shown in Table 1.

The temperature behavior of magnetization, *M(T)* revels important features of the complicated magnetic state of this compound. Figure 2 depicts the temperature dependence of zero field cooled (*ZFC*) and field cooled (*FC*) *DC* magnetization plots for the studied $RuSr_2Eu_{1.4}Ce_{0.6}Cu_2O_{10-\delta}$ (EuRu-1222) sample being measured at 10 *Oe* field. The sharp rise of both *ZFC* and *FC* curve at $T_{mag}$ = 100 K shows a paramagnetic (*PM*) to ferromagnetic (*FM*) transition. The *ZFC* curve shows a peak at around $T_f$ = 85 K, just below the temperature where *ZFC* and *FC* curve separates. At temperature close to $T_{mag}$, the *ZFC* and *FC* curves branch out and the system enters into a so-called spin-glass state, to be discussed later. $T_{mag}$ indicates the onset of the freezing process [20]. The *ZFC* curve has a peak at $T_f$ = 85 K and the *FC* magnetization increases as temperature is decreased. In *ZFC* curve the system undergoes a superconducting transition at $T_c$ = 30 K (marked as the small kink in the *ZFC* curve). Our *DC* magnetization differs from that of the homogeneous spin-glass (SG) behavior. For cluster-spin-glass (*CSG*) the *FC* curve continuous to rise while the same is almost flat for homogeneous spin-glass (*SG*). The increase in the *FC* magnetization below $T_{mag}$ is due to the occurrence of the finite range of inhomogeneous state around a quasi-critical temperature [21, 25]. The magnetic order being induced in rutheno-cuprates stems out from Ru-$O_2$ planes and its details are yet unclear [24, 25]. To further investigate the magnetic nature of the compound at low temperatures, the field dependence of magnetization is performed at 2 K (in magneto-superconducting region). Inset of Figure 2 depicts magnetization (*M*) vs. applied field (*H*) loop in low field range (-500Oe ≤ H ≤ +500Oe) at 2 K. The shape of loop shows a ferromagnetic like behavior, co-existing with the Meissner (diamagnetic) signal at 2 K. The negative moment increases linearly up to 110 Oe, typical for a *SC* state below lower critical field ($H_{c1}$).

Figure 3 shows magnetization (*M*) vs. applied field (*H*) curves of EuRu-1222 sample measured at different temperatures, in the range of -50 kOe to +50 kOe. At low temperatures, the *MH* loops show the *S*-type shape which is the feature of *SG* behavior. With increase in temperature the *S*-type shape is transformed into the linear *PM* shape. The *MH* loop at 200 K is completely a straight line resembling the *PM* state. The *S*-shaped curves also exhibit the coercivity and hysteresis at low fields, indicating the co-existence of *SG* and *FM* states. However, the magnetic saturation is not observed with the applied magnetic field up to the 50 kOe. The magnetization becomes a nonlinear function of field and shows ferromagnetic behavior with magnetic hysteresis in low field region. Both, the absence of magnetization saturation at high- field and the existence of hysteresis loop at low field-regions, are characteristics of spin-glass [29, 30] phase possibly existing with ferromagnetic clusters just below SG state. This will be discussed in later sections with detail. The Arrott plots are also plotted from magnetization vs. field data, see Figure 4. No sign of spontaneous magnetization is seen in Arrott plots instead there is a strong curvature towards the *H/M* axis with no intercept on the $M^2$ axis. The absence of spontaneous magnetization reveals the short range ordering [31], which also indicates towards the possible spin-glass state with ferromagnetic clusters. The situation will be clear, while we discuss the *AC* susceptibility results in next sections.

The main panels of Figures 5(a) and 5(b) show the temperature dependence of the real $\chi'$ (dispersion) and the imaginary part $\chi''$ (absorption) of *AC* susceptibility $\chi_{ac}$, respectively in frequency ranges of 33-9999 *Hz*. Insets of Figures 5(a) and 5(b) show the enlarge view of the *AC* susceptibility of real and imaginary parts respectively. Though the real part c curve exhibit clear peak, the corresponding imaginary part ($\chi''$) curve shows the inflection around the spin-glass ($SG$) transition or freezing point temperature ($T_f$). It can be seen from the inset of Figure 5(a) that the height of the peak around freezing temperature ($T_f$) decreases slightly and is also shifted towards the higher temperature with increasing frequency ($f$). Similarly, for $\chi''$ the height of the peak decreases and is shifted towards higher temperature, see inset Figure 5(b). Though, the qualitative effect is same, the shift with frequency is relatively larger for $\chi''$ than $\chi'$ curve. The increase in the value of $T_f(\chi')$ ($T_f$ = 82.8 K at $f$ = 33 $Hz$ and $T_f$ = 84.4 K at $f$ = 9999 $Hz$) with frequency is the characteristic of spin-glass system and is usually estimated from the quantity $k = \Delta T_f / T_f \Delta(\log_{10} f)$, where Δ refers to the difference in the corresponding quantity. This quantity ($k$) varies in the range of 0.004-0.018 for spin-glass system, while for super-paramagnets it is of the order of 0.3. Taking $T_f$ as the temperature corresponding to maximum value of the $\chi'$ or the inflection point from the $\chi''$ curves, we obtained $k = 7.6 \times 10^{-3}$ or 0.0076, which is in broad agreement with the typical spin-glass system values, e.g., $2.0 \times 10^{-2}$, $1.8 \times 10^{-2}$ and $6 \times 10^{-2}$ for $La(Fe_{1-x}Mn_x)_{11.4}Si_{1.6}$, NiMn and (EuSr)S respectively [32, 33]. It is clear that the studied EuRu-1222 system is a typical spin-glass, with still a possibility of the existence of some inhomogeneous ferromagnetic clusters below $T_f$, to be discussed later. There are basically two different possible interpretations of the spin-glass freezing: the first one assumes the existence of a true equilibrium phase transition at a fixed temperature (canonical-spin-glass) [32]. The second interpretation assumes the existence of ferromagnetic inhomogeneous clusters with non-equilibrium freezing [32]. To further verify the spin-glass (SG) state in EuRu-1222 or for magnetically interacting clusters, the *Vogel-Fulcher law* [17, 25] purposed,

$$\omega = \omega_o \exp.[-E_a/k_B(T_f - T_o)] \quad \ldots\ldots\ldots\ldots\ldots\ldots\ldots\ldots (2)$$

Here $E_a$ is the activation energy or the potential barrier separating two adjacent clusters. $\omega_o$ is the characteristic frequency of the clusters, $T_f$ the freezing temperature and $T_o$ is the *Vogel-Fulcher* temperature, which is the measure of inter-cluster interaction strength. When $T_o$ = 0, *Vogel-Fulcher* law is transformed to the *Arrhenius law* [17, 25], which describes the relaxation processes of non-interacting magnetic clusters as,

$$\omega = \omega_o \exp.[-E_a/k_B T_f] \quad \ldots\ldots\ldots\ldots\ldots\ldots\ldots\ldots\ldots (3)$$

For a magnetic cluster system the $\omega_o/2\pi = 10^{12}$ Hz [16, 25]. The *Vogel-Fulcher* law fits with the experimental data of the EuRu-1222 well (see Figure 6). Figure 6 shows a linear fit between the freezing temperature $T_f$ and $1/\ln(f_o/f)$, the slope of this linear curve gives $E_a/k_B$ = 109.64 K and intercept gives the value of $T_o$ = 78.37 K. The value of the $T_o$ = 78.37 K is in agreement with the value of freezing temperature $T_f$ = 82.8 K, obtained from the *AC* susceptibility measurements. Hence the fitted experimental data of *Vogel-Fulcher* law indicate the presence of spin-glass state in EuRu-1222. For spin-glass (SG) system, the parameter $t^* = (T_f - T_0)/T_f$ must be < 0.10 [25, 26]. In our case the $t^*$ is 0.053 [(82.8-78.37)/82.8], qualifying studied EuRu-1222 to be a spin-glass (SG) system. Worth mentioning is the fact, that we took $\omega_o/2\pi = 10^{12}$ Hz from literature for the same system [16, 27]. Further same characteristic frequency fitted our data reasonably well (Fig.6). Very recently, in case of Fe doped $LaMnO_3$, it is clearly mentioned that choice of characteristic relaxation time $\tau_0$ ($2\pi/\omega_o$) or frequency ($\omega_o/2\pi$) makes a difference in deciding about the presence of SG or cluster-spin-glass (CSG) states [28]. Unfortunately the parameter $t^*$ is not estimated and discussed in earlier reports on similar rutheno-cuprate system [16, 27].

To further investigate the spin-glass state with ferromagnetic clusters, we performed linear and nonlinear *AC* susceptibility measurements on our EuRu-1222 system with varying *AC* drive field and a fixed frequency of 333 *Hz*. Figure 7 (a) and 7 (b) depicts the real ($\chi'_1$) and imaginary parts ($\chi''_1$) of the first harmonic of *AC* susceptibility respectively. Both real ($\chi'_1$) and imaginary part ($\chi''_1$) are measured as a function of temperature from 150 K to 2 K range with zero external *DC* field bias, varying *AC* drive field and a fixed frequency of 333 *Hz*. Figure 7 (a) and (b) show that the peak temperature of both $\chi'_1$ *and* $\chi''_1$ shift slightly towards lower temperature and the height of the peak increases with increasing the amplitude of the *AC* field. This is an *unusual behavior*, because it has been observed earlier that the height of the peak of both $\chi'_1$ *and* $\chi''_1$ decreases as increasing the *AC* field for an ideal SG system [17]. When there is an increase in the amplitude of applied *AC* magnetic field, the magnetic energy associated with the external field becomes large compared to the thermodynamic energy of the magnetic moments. As a result the freezing of magnetic moments does not take place in the direction of the applied field, and hence the magnitude of the $\chi'$ is decreased with increasing applied *AC* drive field. The unusual *AC* drive field dependence of $\chi'$ indicates towards possible presence of embedded ordered magnetic clusters within the *SG* state [17] in studied EuRu-1222 system.

The superconducting transition temperature ($T_c$) can be seen from Figure 7 (a) as a clear change in slope at around 30 K in real ($\chi'_1$) of the first harmonic of *AC* susceptibility at various *AC* drive fields. With further lowering of temperature the diamagnetic signal is also seen below 20 K. The superconducting state is superimposed with positive background magnetism of the system and hence the diamagnetism is seen 15 K below the exact $T_c$ of around 30 K.

Finally further confirm the presence of ferromagnetic clusters within *SG* state, we also measured the field dependence of magnetization (*MH*) at different temperatures (50, 75, 100 and 125 K) in low field (< 6 kOe) regime. It is clear that EuRu-1222 is paramagnetic at 200 K, S type SG at 100 K and at further lower temperatures of 75 K and 50 K shows a clear ferromagnetic (*FM*) opening. While the magnetic saturation do not takes place at applied field up to 50 kOe (figure 3). This confirms the presence of ferromagnetic clusters in the studies system followed by spin-glass state.

As we have already discussed above from both *AC* (*AC* drive field dependence) and *DC* magnetization (*MH*), the spin-glass and ferromagnetic clusters co-exist in EuRu-1222 system. The even harmonic ($\chi'_2$) signal exhibits a sharp negative peak around the *SG* freezing transition temperature of 80 K along with a broadened shoulder at close to cluster forming temperature of 60 K, see Figure 8. More clearly, Figure 8 exhibits the real parts of the second harmonic ($\chi'_2$) and the third harmonic ($\chi'_3$) of *AC* susceptibility as a function of temperature measured at *AC* drive field of 10 Oe and frequency 333 Hz. The second harmonic of *AC* susceptibility ($\chi'_2$) originates a negative peak near the peak temperature of first harmonic of the $\chi'_1$ (Figure 5a) i.e., at spin-glass transition/freezing temperature of *FM* clusters. The magnetization is represented by equation (1). But if there is a direct *PM* to *SG* type transition, magnetization (*M*) can be expressed as an odd power series of *H* [25].

$$M = M_0 + \chi_1 H + \chi_3 H^3 + \chi_5 H^5 + \ldots\ldots\ldots\ldots \quad (4)$$

Where, $\boldsymbol{\chi_1}$, $\boldsymbol{\chi_3}$ and $\boldsymbol{\chi_5}$ are the odd harmonics of the *AC* susceptibility. Odd harmonics are more important than the even ones to probe complex magnetic systems. It is seen from Figure 8, that $\chi'_3$ shows a positive peak near the temperature where $\chi'_2$ shows a negative peak corresponding to the *SG* transition. This type of behavior is observed in other *CG* systems [35, 36]. It is also observed that $\chi'_2$ shows a small peak at 110 K as well, which resembles the magnetic ordering temperature $T_m$ (*DC ZFC-FC* branching, Figure 2). Hence it is clear from Figures 5 and 7 that the *SG* transition and formation of non homogenous *FM* clusters take place below the magnetic ordering temperature ($T_m$). Further a small negative peak near to the main positive peak in $\chi'_3$ and the change in the sign of $\chi'_2$ from negative to positive during the temperature interval 100 K ≤ *T* ≤ 77 K are some other unusual feature observed in Figure 8.

These may be related to some other minor magnetic transformations taking place in EuRu-1222, which are not clear to us at this moment.

The temperature dependence of the electrical resistivity *ρ(T)* in various applied fields is shown in Figure 9. In order to probe the effect of magnetic field on the superconducting properties of this Ru based compound, we carried out magneto-resistivity measurements in fields up to 130 kOe. Superconducting transition temperature ($T_c$) onset and $\rho$ = 0 are seen at around 30 K and 18 K respectively without any applied field and the same decreases to 10 K and 2 K respectively under 130 kOe field. As reported by Tokunaga and co-authors [37], it is possible that the broadening of the superconducting transition could be due to the presence and motion of self induced vortices. This kind of behavior is frequently observed in granular bulk superconductors. They found a broad resistive transition in $RuSr_2YCu_2O_8$, a material with no magnetic rare earth ions, was correlated with the occurrence of the development of spontaneous moments within the Ru-$O_2$ planes. Increasing fields result in appreciable changes in the shape of *ρ (T)* in the superconducting state. As the magnetic field increases, the superconducting transition becomes broader near the onset of superconductivity, and sharper near the zero resistance state. Inset of Figure 9 shows the normal state resistivity ρ (*T*) plot between 5 K and 200 K for zero applied fields. As we move from higher temperature to the lower the resistivity first decreases slightly exhibiting a metallic behavior. This behavior continues until the temperature reaches a minimum at a *SG* freezing temperature $T_f \approx 83$ K, after that resistivity starts to increase, and finally drops sharply at *SC* transition $T_c$ (onset) = 30 K. In EuRu-1222, the increase of resistivity below $T_f$ is in contrast to the one reported for canonical-spin-glasses, such as AuCr, CuMn, having a resistivity maximum at $T_f$ [25]. These systems are metallic spin glasses, with diluted magnetic impurities, dominated by Ruderman-Kittel-Kasuya-Yoshida (RKKY) interaction between impurity spins [38]. A similar resistivity minimum near spin glass transition has been observed in a number of *d*- and *f*-electron spin-glasses [39, 40], in NiMn [41], NiMnPt alloys [42] and atom disordered uranium compound $U_2IrSi_3$ [43]. In NiMn and NiMnPt alloys, the minimum in resistivity was due to the mixed ferromagnetic and spin-glass state.

The minimum in the *ρ (T)* curve of the EuRu-1222 mimics the metal to insulator transition. Further lowering the temperature, when the material is in near insulating state, the superconductivity occurs at $T_c$. This type of behavior is also observed for high temperature cuprate superconductors, where the ground state is antiferromagnetic *AFM* insulators and superconductivity sets on the injection of charge carriers means superconductivity occurs in the vicinity of the metal-insulator transition. The over-doped range of the material exhibits the normal metal like behavior while the under-doped samples show the insulator-like behavior. It has been observed that in the under-doped systems, the superconducting state co-exist with the spin-glass nature throughout the bulk sample [44]. At low temperature, where the spin-glass phase dominates, the spin freezing affect the motion of the charge carriers and lowers the charge mobility by inducing disorder in the system [44]. Hence in EuRu-1222 system, the formation of inhomogeneous *FM* clusters creates the glassiness and disorder, which play a significant role to the metal-insulator transition.

## Conclusion

This EuRu-1222 system shows a rich variety of magnetic phenomena. A *PM* to *FM* transition at around 110 K, Spin-glass transition temperature ($T_f$) around 82 K and formation of non-homogeneous *FM* clusters at further lower temperature. The *DC* and *AC* susceptibility studies presented in this paper clearly reveal that a *SG* state exists with non-homogeneous *FM* clusters. The frequency dependent peak observed in the $\chi_{ac}$ provides strong evidence of the important role of magnetic frustration in EuRu-1222 and spin-glass properties over a particular temperature range. The *FM* order and spin-glass state co-exist below freezing temperature ($T_f$). Superconductivity is seemingly not affected by various co-existing magnetic phenomena. In last, our results support the presence of spin-glass state with non-homogeneous ferromagnetic clusters in EuRu-1222 system. The oxygen non-stoichiometry in EuRu-1222 might be the reason. Possible random distribution of $Ru^{5+}$-$Ru^{5+}$, $Ru^{4+}$-$Ru^{5+}$ and $Ru^{4+}$-$Ru^{4+}$ exchange interactions may be responsible for observed *SG* with ferromagnetic clusters.

## Acknowledgements

The authors from NPL would like to thank Prof. R. C. Budhani (Director, NPL) for his keen interest in the present work. Authors would like to thank Prof. I. Felner from Hebrew University Jerusalem for careful reading of our MS and various fruitful discussions. One of us, Anuj Kumar, would also thank Council of Scientific and Industrial Research (*CSIR*), New Delhi for financial support through Senior Research Fellowship (*SRF*). This work is also financially supported by *DST-SERC* (Department of Science and Technology), New Delhi funded project on Investigation of pure and substituted Rutheno-cuprate magneto-superconductors in bulk and thin film form at low temperature and high magnetic field.

# Figure Captions

**Fig. 1** Observed (*solids circles*) and calculated (*solid lines*) *XRD* patterns of $RuSr_2Eu_{1.4}Ce_{0.6}Cu_2O_{10-\delta}$ compound at room temperature. *Solid lines* at the bottom are the difference between the observed and calculated patterns. *Vertical lines* at the bottom show the position of allowed Bragg peaks.

**Fig. 2** *ZFC* and *FC DC* magnetization plots for $RuSr_2Eu_{1.4}Ce_{0.6}Cu_2O_{10-\delta}$ measured in the applied magnetic field, $H = 10$ Oe. Inset shows the *M* vs. *H* plot at temperature 2K in the range of -500 Oe $\leq$ H $\leq$ +500 Oe.

**Fig. 3** Typical magnetization loops as function of applied magnetic field measured at different temperatures in the range - 50 kOe to + 50 kOe.

**Fig. 4** Arrott plots ($H/M$ vs. $M^2$) using *DC* magnetization vs. applied field data observed at different fixed temperatures (5, 20, 50, 75, 100 and 125 K).

**Fig. 5(a)** Temperature dependence of the real part of *AC* susceptibility, measured at different frequency with zero external *DC* magnetic field. Inset shows the enlarged view of the real part of the first harmonic *AC* susceptibility. The cusp shown in real part of *AC* susceptibility is corresponding to the inflection point in the imaginary part of *AC* susceptibility.

**Fig. 5(b)** Temperature dependence of the imaginary part of *AC* susceptibility, measured at different frequency with zero external *DC* magnetic field. Inset shows the enlarged view of the imaginary part of the first harmonic *AC* susceptibility.

**Fig. 6** The variation of the freezing temperature $T_f$ with the frequency of the *AC* field in a *Vogel-Fulcher* plot. The solid line is the best fit of equation (2).

**Fig. 7(a)** Temperature dependence of the real part of *AC* susceptibility measured at different amplitude with zero external *DC* magnetic fields. The cusp shown in real part of *AC* susceptibility is corresponding to the inflection point in the imaginary part of *AC* susceptibility.

**Fig. 7(b)** Temperature dependence of the imaginary part of *AC* susceptibility, measured at different amplitude with zero external *DC* magnetic field.

**Fig. 7(c)** Low field magnetization loops as function of applied magnetic field measured at different temperatures (50, 75, 100, 125 and 200K) in the range - 6000 Oe to + 6000 Oe.

**Fig. 8** The real parts of the second harmonic ($\chi'_2$) and the third harmonic ($\chi'_3$) of *AC* susceptibility as a function of temperature measured at *AC* field 10 Oe and frequency 333 Hz.

**Fig. 9** Resistivity vs. Temperature ($\rho$-T) plot for $O_2$-annealed $RuSr_2Eu_{1.4}Ce_{0.6}Cu_2O_{10-\delta}$ at various applied magnetic field. Inset shows the normal state resistivity $\rho$ (T) as a function of temperature in zero fields.

**Table 1** Atomic coordinates and site occupancy for studied $RuSr_2Eu_{1.4}Ce_{0.6}Cu_2O_{10-\delta}$

Space group: ***I4/mmm***, Lattice parameters; ***a* = 3.8384 (3) Å, *c* = 28.5769 (2) Å, $\chi^2$ = 3.33**

| **Atom** | **Site** | ***x* (Å)** | ***y* (Å)** | ***z* (Å)** |
|---|---|---|---|---|
| **Ru** | *2b* | 0.0000 | 0.0000 | 0.0000 |
| **Sr** | *2h* | 0.0000 | 0.0000 | 0.4221 (2) |
| **Eu/Ce** | *1c* | 0.0000 | 0.0000 | 0.2948 (4) |
| **Cu** | *4e* | 0.0000 | 0.0000 | 0.1443 (3) |
| **O(1)** | *8j* | 0.6057 (4) | 0.5000 | 0.0000 |
| **O(2)** | *4e* | 0.0000 | 0.0000 | 0.0594 (4) |
| **O(3)** | *8g* | 0.0000 | 0.5000 | 0.1445 (5) |
| **O(4)** | *4d* | 0.0000 | 0.5000 | 0.25000 |

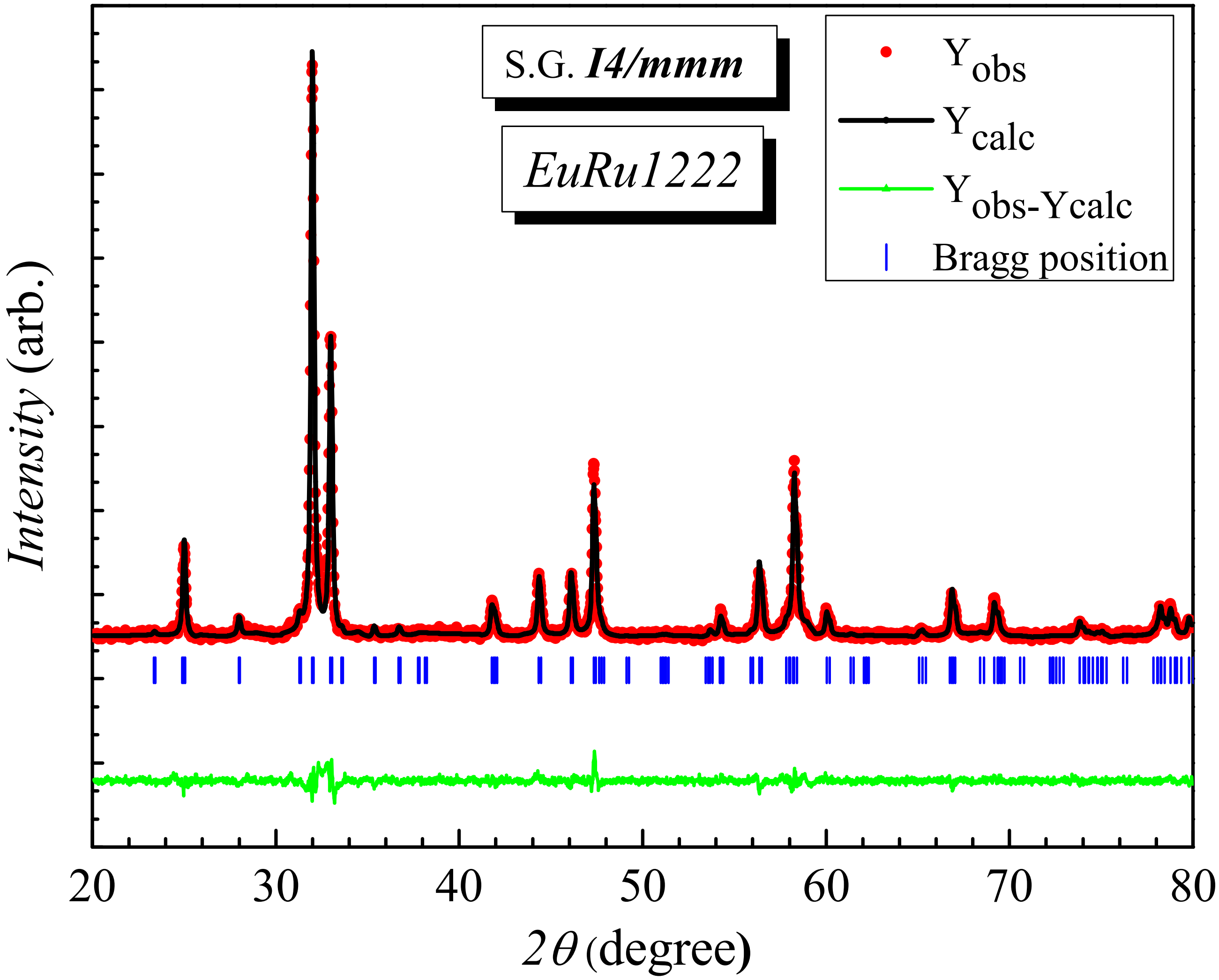

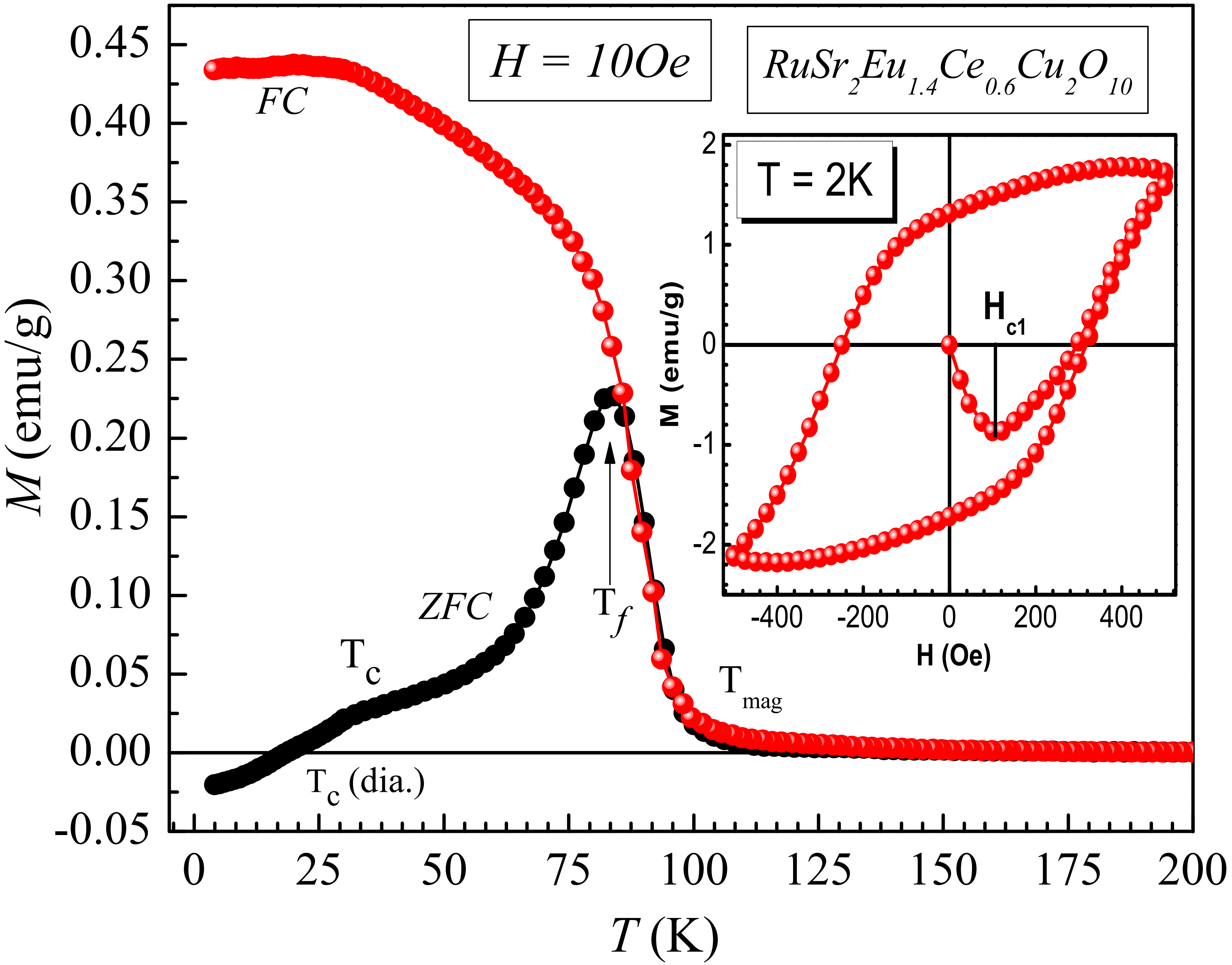

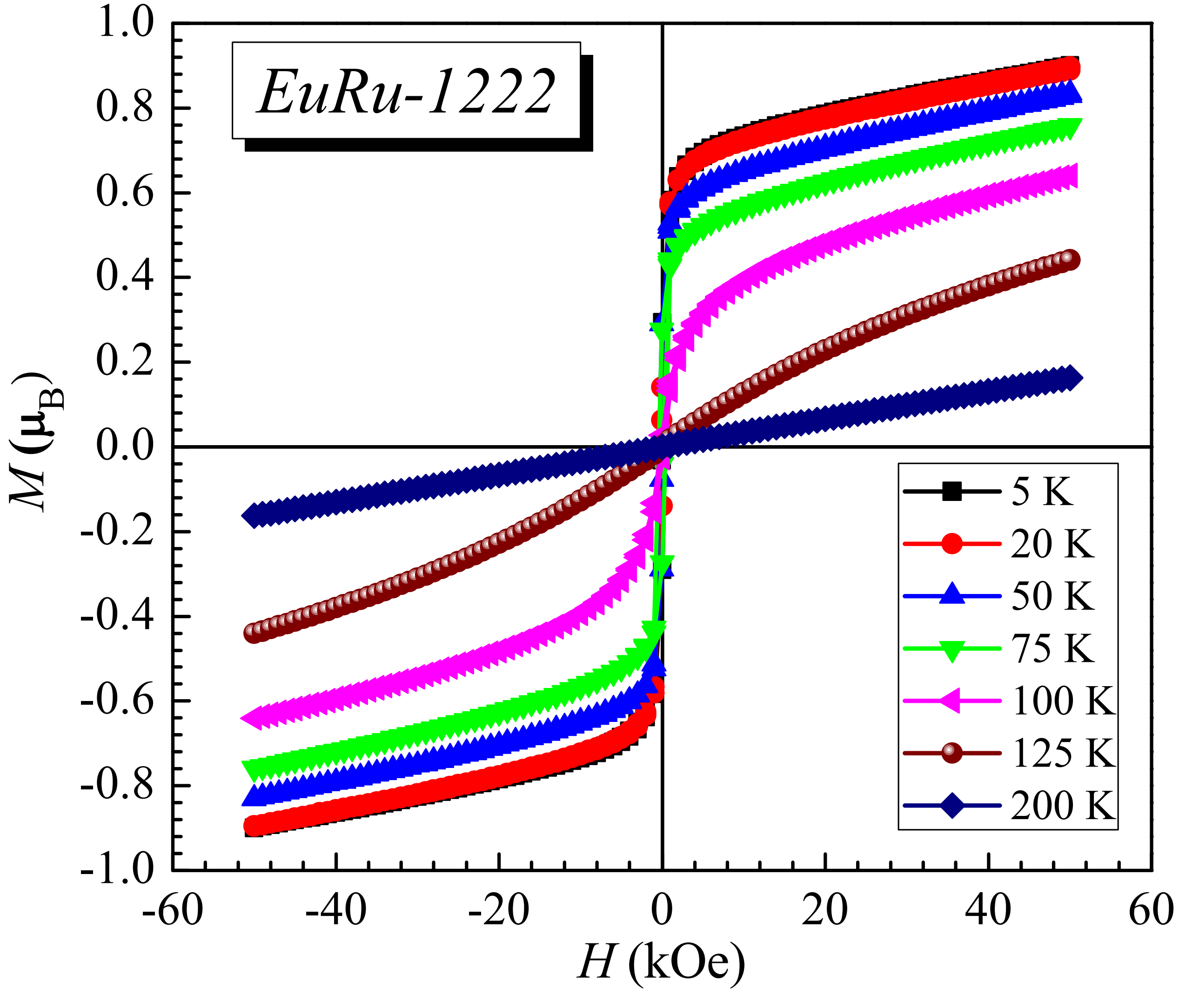

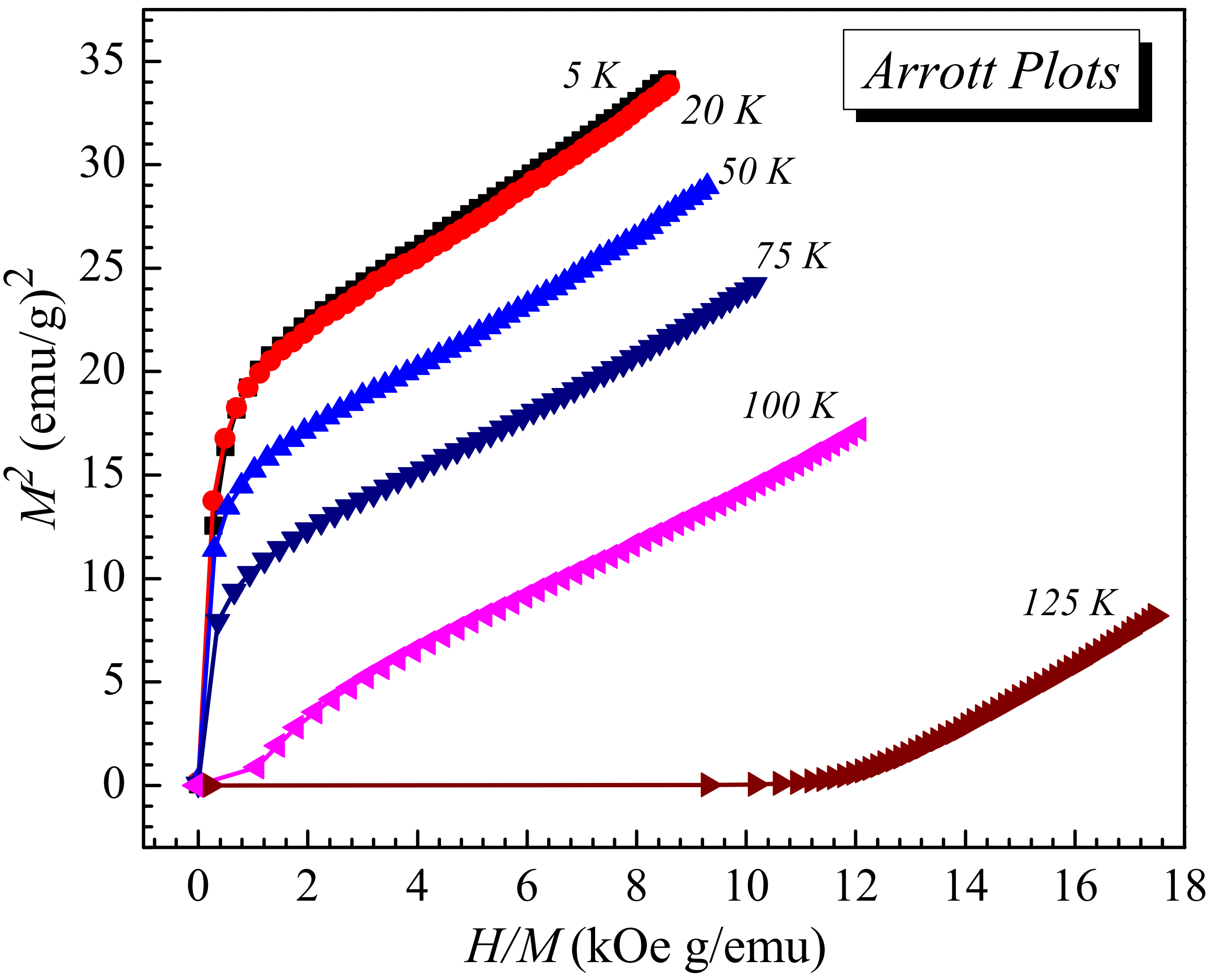

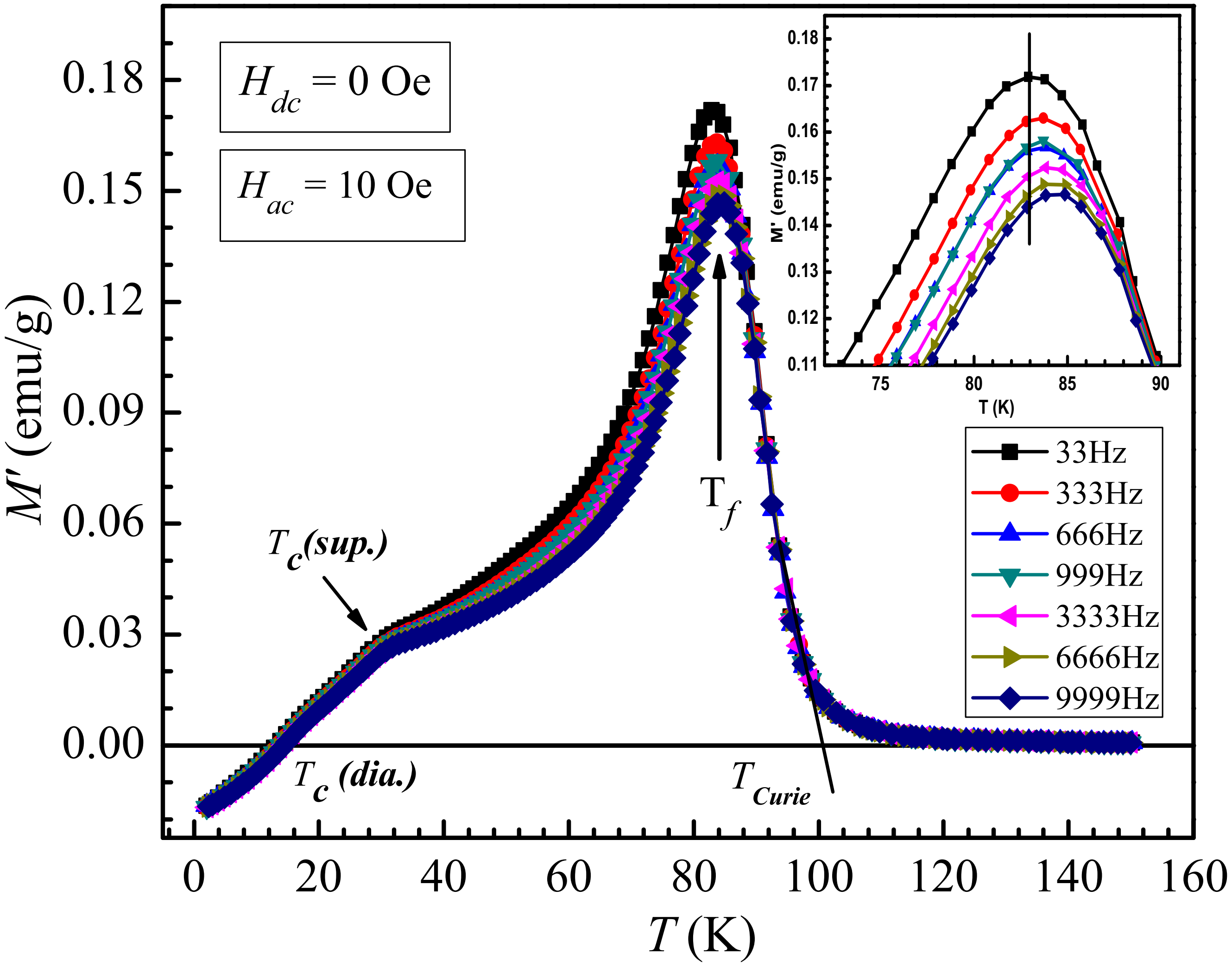

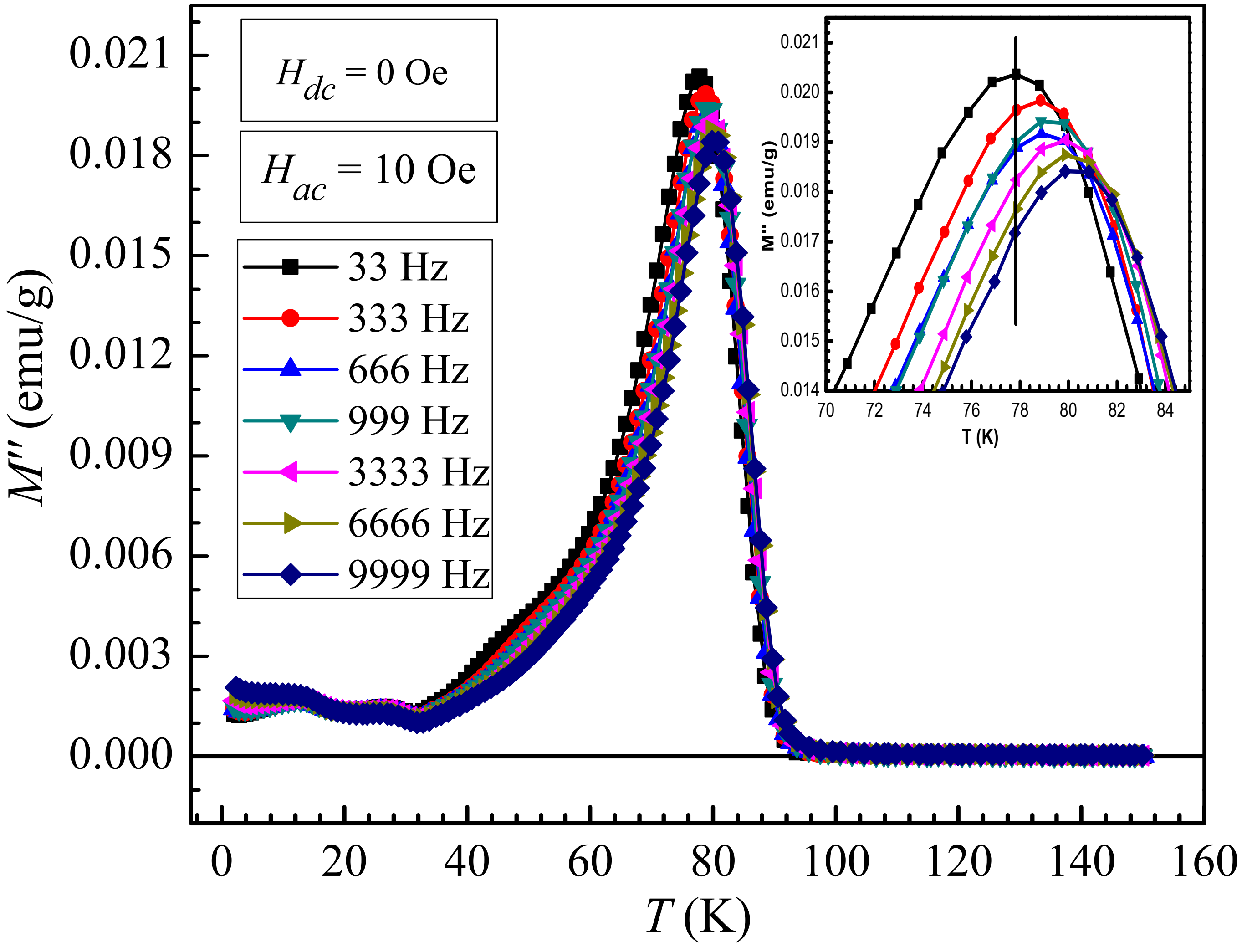

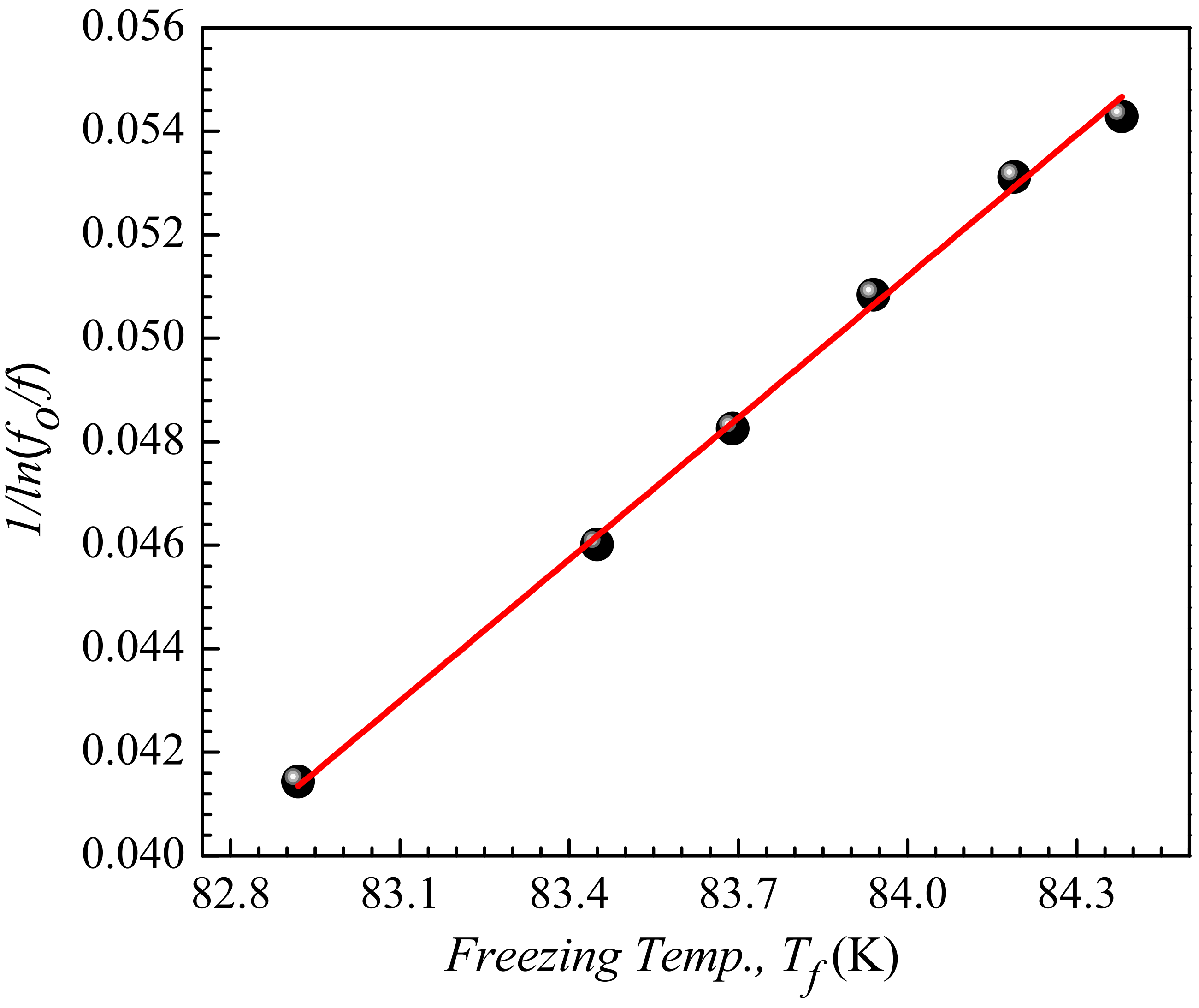

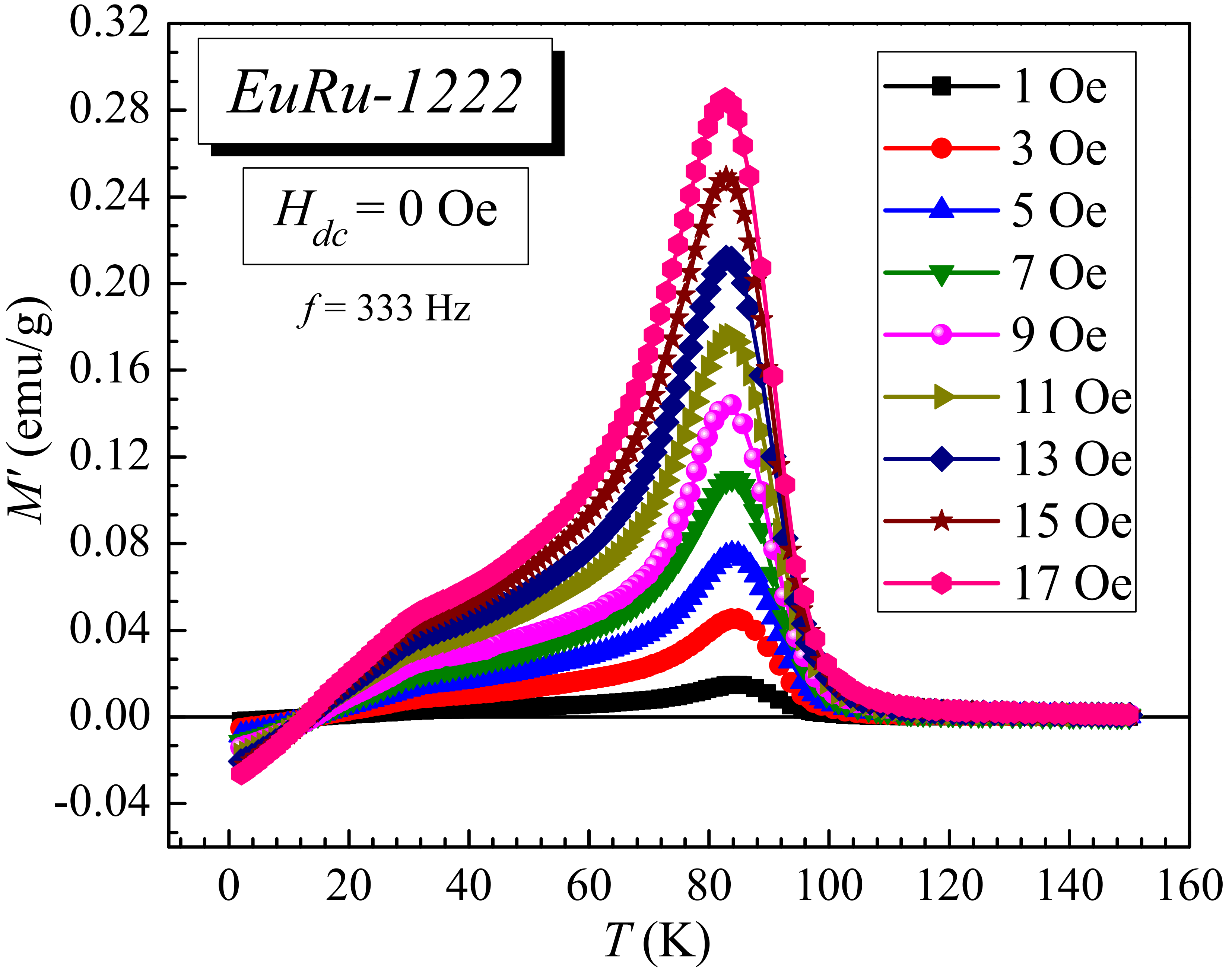

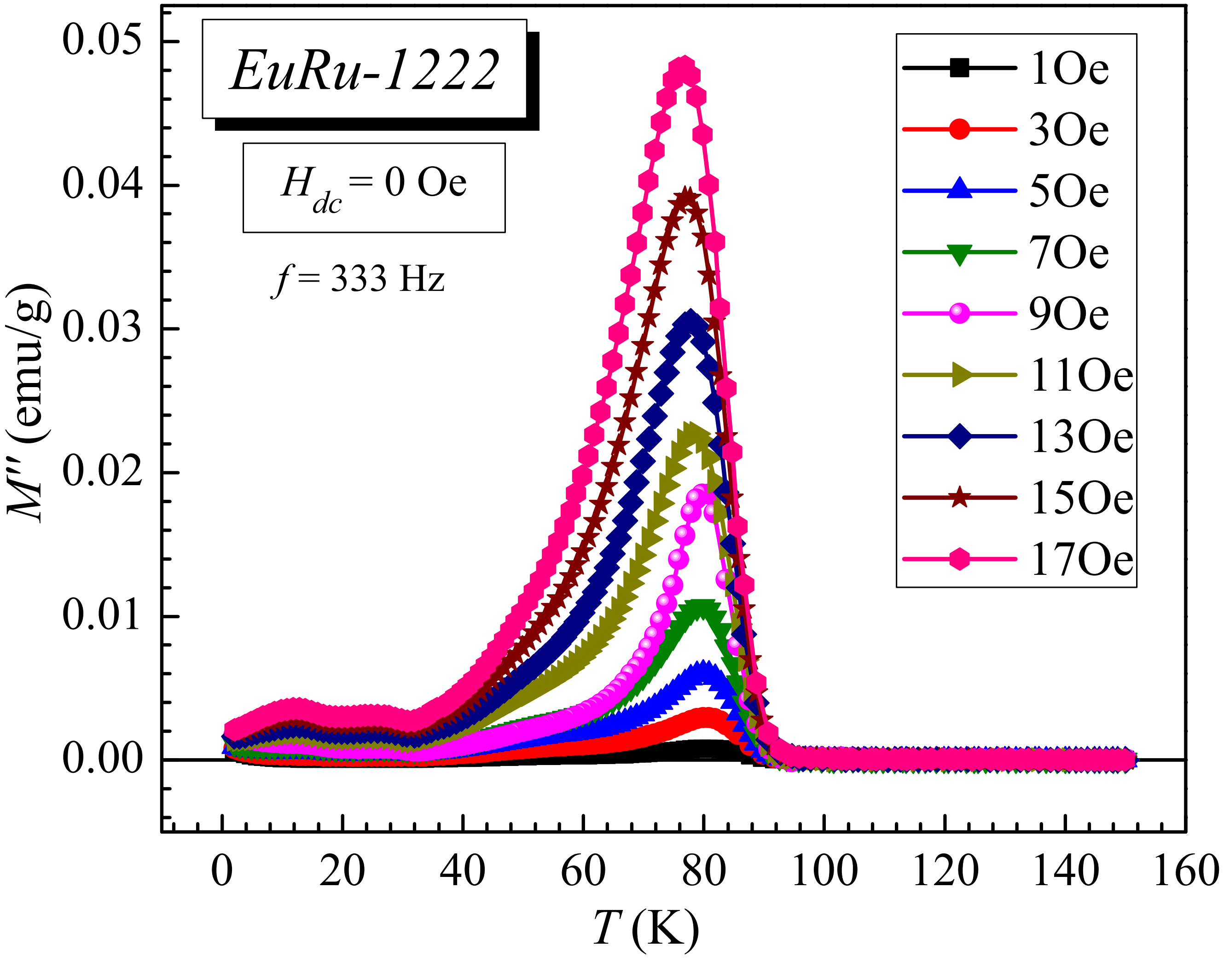

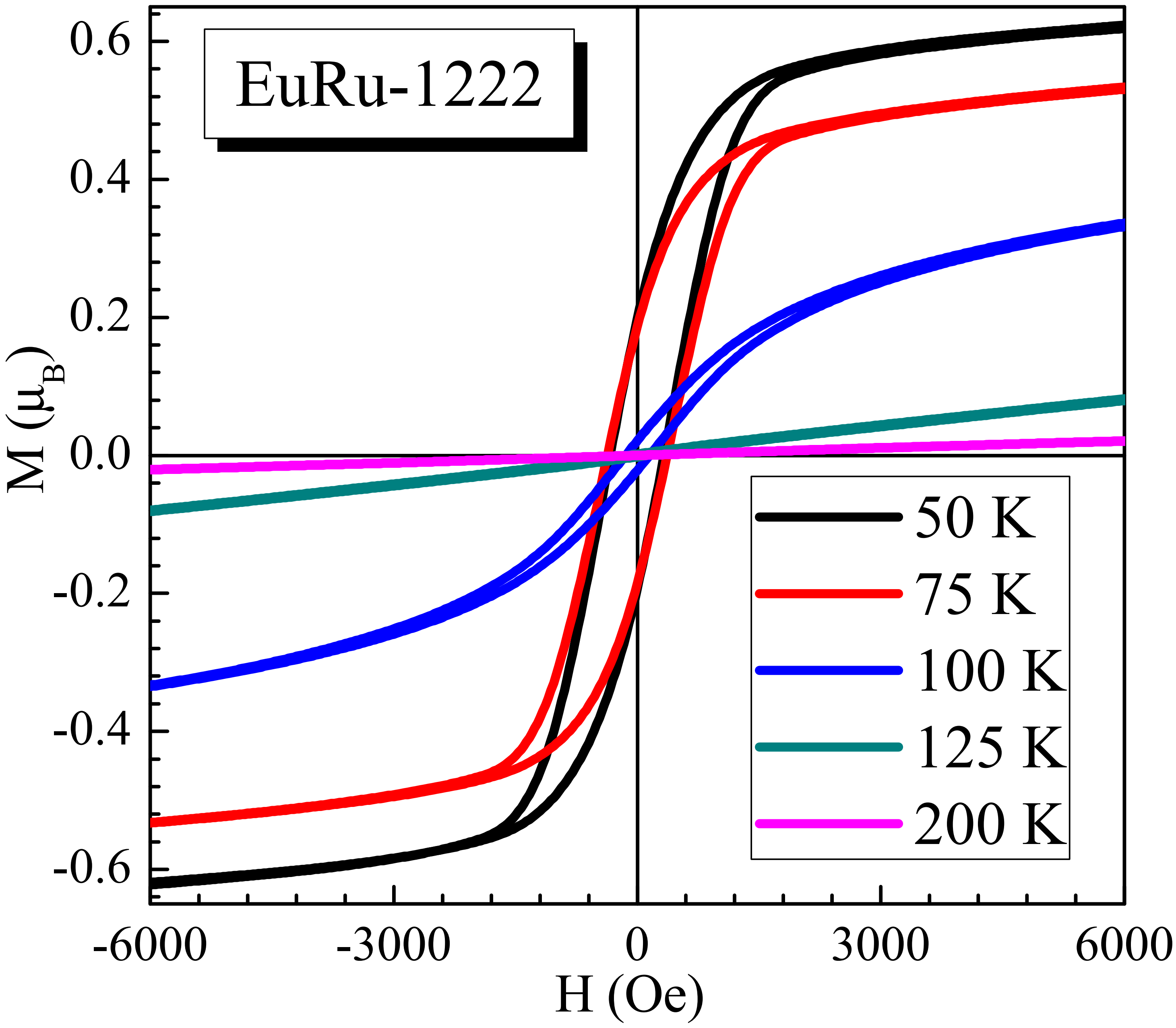

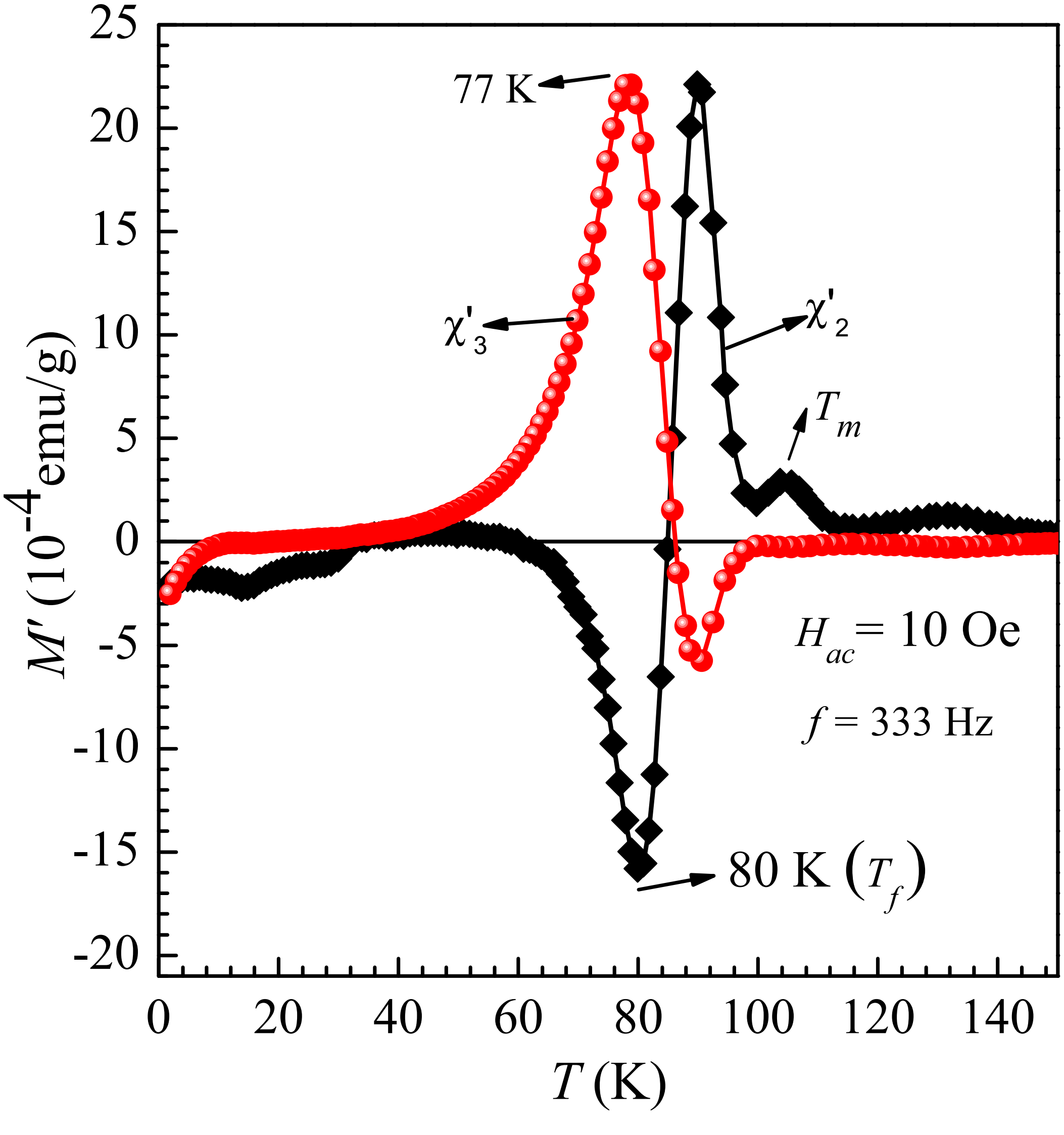

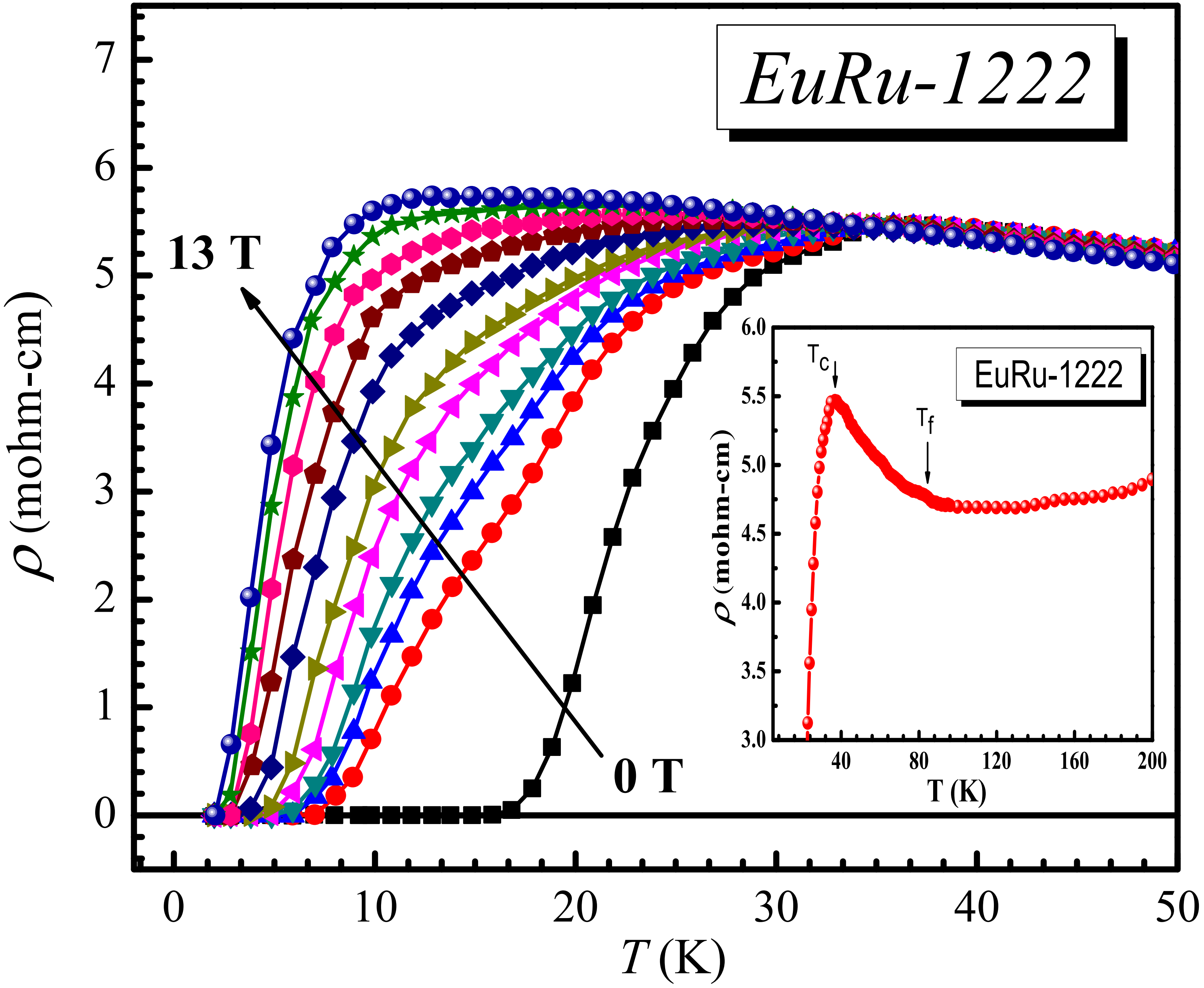